\documentclass[conference,10pt]{IEEEtran}
\usepackage{graphicx}
\usepackage{amssymb,amsmath}
\usepackage{epstopdf}
\usepackage{amsthm}
\usepackage{color}
\usepackage{amsfonts}
\usepackage{amsfonts}
\usepackage{cite}
\usepackage{algorithmic}
\usepackage[center]{caption}
\usepackage{epsfig}
\usepackage{latexsym}
\usepackage{epstopdf}
\usepackage{verbatim}
\usepackage{amsthm}
\usepackage[linesnumbered,ruled,vlined]{algorithm2e}

\graphicspath{{./figures/}}
\IEEEoverridecommandlockouts

\begin{document}

\title{A DoF-Optimal Scheme for the two-user X-channel with Synergistic Alternating CSIT}

\author{\large Ahmed Wagdy$^\dagger$, Amr El-Keyi$^\dagger$, Tamer Khattab$^\star$, Mohammed Nafie$^\dagger$ \\ [.1in]
\small
\begin{tabular}{c} $^\dagger$Wireless Intelligent
Networks Center (WINC), Nile University, Smart Village, Egypt.\\

$^\star$Electrical Engineering, Qatar University, Doha, Qatar.\\

\thanks{This work was supported by the Qatar National Research
Fund (QNRF) under Grant NPRP 6-1326-2-532.}\end{tabular}
}


\maketitle

\begin{abstract}
In this paper, the degrees of freedom (DoF) of the two-user single input 
single output (SISO) X-channel are investigated. Three
cases are considered for the availability of channel state
information at the transmitters (CSIT); perfect, delayed, and no-CSIT.
A new achievable scheme is proposed to
elucidate the potency of interference creation-resurrection (IRC) when the available CSIT
alternates between these three cases. For some patterns of
alternating CSIT, the proposed scheme achieves $4/3$ DoF, and
hence, coincides with the information theoretic upper bound on the DoF of the
two-user X-channel with perfect and instantaneous CSIT. The
CSIT alternation patterns are investigated where the patterns that 
provide extraordinary synergistic gain and dissociative ones are identified. 
\end{abstract}
\vspace{2 mm}

\IEEEpeerreviewmaketitle

\vspace{2mm}
\section{Introduction}

Breaking the interference barrier is an important step towards
unleashing the capacity of wireless networks to meet future
demands. Several classical interference management approaches have
been utilized such as successive interference cancellation,
treating the interference as noise, and orthogonalizing the
channel access. Interference alignment (IA) is an interference
management technique that refers to creating a correlation or an
overlap between the interference signals at the receiver in order
to minimize dimensions of the interference subspace and maximize
the desired signal space dimensions \cite{jafar2011interference
,maddah2008communication,4567443}. Global, perfect, and
instantaneous channel state information at the transmitters (CSIT)
is an essential component for interference alignment in pioneering
works \cite{maddah2008communication,4567443,5208535,4418479}. On
the other hand, in the complete absence of CSIT, an overly
pessimistic assumption, the potential of IA fades and the degrees
of freedom (DoF) of many networks collapse to what is achievable
by time-division multiplexing \cite{6205390}.

Contrary to the popular belief that in fast fading environments
delayed CSIT is vain, Maddah Ali and Tse
\cite{maddah2012completely} proved that completely outdated CSIT
can still be useful even if the channel states are completely
independent. Various interesting extensions of
\cite{maddah2012completely} have been considered, e.g., the 3-user
interference channel 
\cite{maleki2012retrospective} and the two-user MISO broadcast channel
with imperfect instantaneous CSIT and perfect delayed CSIT
\cite{yang2012degrees}. 

In another research direction, Tandon et al. in \cite{6471826}
formalized an interesting model for the availability of the CSIT,
called alternating CSIT, which provides practical standpoints. In
this model, the authors allow the availability of CSIT to vary
over time which is a more practical assumption and convenient to
the nature of wireless networks. 
They proved that alternating CSIT could be beneficial. For example, in
the case of a two-user MISO broadcast channel with perfect CSIT
for one user and no-CSIT for the other user, the authors
conjectured that the optimal DoF with fixed CSIT is only one while
in the alternating CSIT setting $3/2$ DoF can be achieved.

Earlier research work on the DoF of the two-user X-channel have
determined that the upper bound for DoF of two-user single-input
single-output (SISO) X-channel is $4/3$ 
\cite{maddah2008communication, 5208535,4418479}. This upper
bound is achievable with global, perfect, and instantaneous CSIT
and when the channel coefficients are time varying. 
The authors of
\cite{maleki2012retrospective} showed that even in a fast fading
environment and for interference networks consisting of
distributed transmitters and receivers, delayed CSIT could
be beneficial and have a great impact on increasing the DoF. They
proved that for the two-user SISO X-channel, $8/7$ DoF is
achievable with delayed CSIT. New results have been demonstrated
in \cite{ghasemi2011degrees} where the two user SISO X-channel
with delayed CSIT could achieve $6/5$ DoF.

In this work, we consider the two-user SISO X-channel with
alternating CSIT. The main question we ask is whether the
\textit{synergistic alternation} in the availability of CSIT  is
beneficial in this channel as it is in collocated transmitters
networks. We answer this question in an affirmative way by
presenting optimal scheme that exploits the synergistic
alternation of the CSIT to achieve the upper bound on the DoF of
the two-user SISO X-channel.

The second question we ask is whether all alteration patterns  have
synergistic benefits. We negatively answer this question and show
that there exists some certain alteration patterns in which some
channel knowledge availability states can not work together in a
cooperative way but they work individually and the corresponding
DoF dwindle to the sum of their individual DoF. Furthermore we
find the synergistic alternation patterns and dissociative ones.

\section{System Model}
A two-user SISO X-channel is considered in which two
transmitters $T_1$ and $T_2$  transmit four independent messages
$W_{11}, W_{12},W_{21}, W_{22}$ to receivers $R_1$ and $R_2$,
where $W_{ij}$ originates in transmitter $j$ and is intended to
receiver $i$ and each node equipped with single antenna. The
received signals at the $i$th receiver at time slot $t$ is given
by
\begin{equation}\label{eq1}
Y_i(t)= \displaystyle\sum_{j=1}^2 h_{ij}(t)X_j(t)+N_i(t)
\end{equation}
where $X_j(t)=f_{1j}(t) W_{1j}(t)+f_{2j}(t) W_{2j}(t)$ is the
transmitted signal from $T_j$ at the $t$th time slot which
satisfies the power constraint $E[\|X(t)\|^2]\leq P$ and
$f_{ij}(t)$ is the precoding coefficient for the message
$W_{ij}(t)$. The noise $N_i(t)\sim \mathbb{C}\mathcal{N}(0,1)$ is
the circularly symmetric white Gaussian noise with unit variance
generated at $R_i$ at time slot $t$. In \eqref{eq1}, $h_{ij}(t)$
is the channel coefficient from $T_j$ to $R_i$ and all channel
coefficients are independent identically distributed and drawn
from continuous distribution. Let $r_{ij}(P)$ denote the
achievable rate of $W_{ij}$ for a given transmission power $P$
where $ r_{ij}(P)= \frac{\log_2(|W_{ij}|)}{n}$ and $n$ is the
number of channel uses. We are interested in the DoF region
$\mathcal{D}$, defined as the set of all achievable tuples
$(d_{11}, d_{12}, d_{21},d_{22})$, where $d_{ij}= \lim_{P
\rightarrow \infty} \frac{r_{ij}(P)}{\log_2{(P)}}$ is the DoF for
message $W_{ij}$. The total DoF of the network is defined as
\begin{equation}
\text{DoF}= \max_{(d_{11},d_{12},d_{21},d_{22}) \in  \mathcal{D}}
d_{11}+d_{12}+d_{21}+d_{22}
\end{equation}

We assume that the receivers have perfect global channel state
information. Furthermore, we consider three different states of
the availability of CSIT identified by:
\begin{enumerate}
  \item Perfect CSIT (P): identifies the state of CSIT in which CSIT is available to the transmitters instantaneously and without error.
  \item Delayed CSIT (D): identifies the state of CSIT in which CSIT is available to the transmitters with some delay $\ge$ one time slot and without error.
  \item No CSIT (N): identifies the state of CSIT in which CSIT is not available to the transmitters at all.
\end{enumerate}
The state of CSIT availability of the channels to the $i$th
receiver is denoted by $S_i$; where, $S_i$ $\in \{\text{P,D,N}\}$.
In addition, let $S_{12}$ denote the state of CSIT availability
for the channels to the first and second receivers, respectively.
Therefore, $S_{12} \in \{\text{PP,PD,PN,DP,DD,DN,NP,ND,NN}\}$. For
example, $S_{12}= \text{PN}$ refers to the case where $T_1$ has
perfect knowledge of $h_{11}$ (and no information about $h_{21}$)
and $T_2$ has perfect knowledge of $h_{12}$ (and no information
about $h_{22}$).

We denote the CSIT availability of the channels to the $i$th
receiver over three time slots by $S_i^T=(x,y,z)$ where $x,y,z \in
S_{i}$ and $x, y, \text{and } z$ denote the availability of the
CSIT in the first, second, and third time slots respectively.
Similarly, we denote the availability of CSIT for the channels to
the first and second receivers in three time slots \textquotedblleft CSIT pattern\textquotedblright denoted by
$S_{12}^T=(X,Y,Z)$ where $X,Y.Z \in S_{ij}$.

\section{Proposed Achievable Scheme} \label{section_schems}
Motivated by the fact that multiuser networks with time varying channels, 
the variation in the availability of CSIT or
the fluctuation in state of CSIT across different links is
inevitable, we extend this verity modelled for MISO broadcast
channel in \cite{6471826} to the two-user SISO X-channel. Form
\cite{maddah2008communication, 5208535,4418479}, it is known that
for the two-user SISO X-channel the DoF of the network are bounded
by $4/3$. This upper bound is achievable over $3$-symbol channel
extension if the channel coefficients vary over time 
and each transmitter has global and \textit{constantly} perfect
CSIT over the $3$ time slots.

In this section, we present three illustrative examples for the
proposed achievable scheme in three different patterns of CSIT
availability. In all these cases, we show that $4/3$ DoF is
achievable by sending $4$ different data symbols; $2$ for each
receiver over three time slots. The basic idea behind the proposed
achievable scheme is to resurrect the interference formerly created, 
hence, interference creation-resurrection (ICR)
strikes and  interference alignment arises.
Inspired by the MAT
algorithm in \cite{maddah2012completely}, the proposed achievable
scheme is performed in two phases over three time slots.

The first phase is associated with the delayed CSIT state where
the transmitters transmit their messages. As a result, the
receivers get linear combinations of their desired messages in
addition to interference. This phase is called ``interference
creation'' phase. On the other hand, the second phase is
associated with the perfect CSIT state and is called the
``interference resurrection'' phase. In this phase, the
transmitters reconstruct the old interference by exploiting the
delayed CSIT received in phase one and the perfect CSIT in the
second phase. Hence, after three time slots, each receiver has two
different linear combinations of its desired messages and only one
interference term received twice. Noteworthy, in some cases the
two phases can overlap over the 3 time slots.

Let $u_1$ and $u_2$ be two independent symbols intended to $R_1$
transmitted from $T_1$ and $T_2$, respectively. Also, let $v_1$
and $v_2$ be two independent symbols intended to $R_2$ from $T_1$
and $T_2$, respectively. In the next subsections, we show that we
can reliably transmit these  symbols to their target destinations
in 3 time slots in three different cases of alternating CSIT.

\subsection{Scheme 1: Combined delayed and distributed perfect CSIT}
As an illustrative example of this case, let us consider a 2-user
SISO X-channel with alternating CSIT  pattern given by (DD, PN,
NP) over three time slots. Here, we have combined delayed CSIT in
the first time slot and distributed perfect CSIT over the last two
time slots. Consequently, the proposed scheme is performed in two
separate phases as follows.

\textit{Phase one:} In this phase, the two transmitters greedily
transmit all data symbols, i.e., $X_1(1)= u_1 + v_1$ and $X_2(1)=
u_2 + v_2$. As a result, the received signals are given as
\begin{eqnarray}
Y_1(1)&\!\!\!=\!\!\!& h_{11}(1)u_1 + h_{12}(1)u_2+ h_{11}(1)v_1+h_{12}(1)v_2 \nonumber \\
&\!\!\!\equiv \!\!\!& L_1^1(u_1,u_2)+ I_1(v_1,v_2)\\
Y_2(1)&\!\!\!=\!\!\!&h_{21}(1)u_1 + h_{22}(1)u_2+ h_{21}(1)v_1+h_{22}(1)v_2 \nonumber \\
&\!\!\!\equiv\!\!\!& I_2(u_1,u_2)+ L_2^1(v_1,v_2)
\end{eqnarray}
where $L_i^j(x_1,x_2)$ denotes the $j$th linear combination of the
two messages $x_1$ and $x_2$ that are intended for receiver $R_i$
and $I_i(z_1,z_2)$ denotes the interference term for
receiver $R_i$ which is a function of the messages $z_1$ and $z_2$
that are not intended for this receiver.

\textit{Phase two:} This phase consists of two time slots
where in each time slot the transmitted signals are designed such
that the interference is resurrected at one receiver while the
second receiver receives a new linear combination of its desired
messages. Note that now the transmitters are aware of the CSIT of
the previous time slot, i.e., $T_1$ knows $h_{11}(1)$ and
$h_{21}(1)$ while $T_2$ knows $h_{12}(1)$ and $h_{22}(1)$. Also,
at $t=2$, the channels to the first receiver are known perfectly
and instantaneously at the two transmitters, i.e., $T_1$ knows
$h_{11}(2)$ and $T_2$ knows $h_{12}(2)$. As a result, the first
time slot in this phase is dedicated to resurrecting the
interference $I_1(v_1,v_2)$ received by $R_1$ in the first time
slot. The transmitted signals of $T_1$ and $T_2$ are given by
\begin{eqnarray}
X_1(2)&\!\!\!=\!\!\!&h_{11}^{-1}(2)h_{11}(1)v_1\\
X_2(2)&\!\!\!=\!\!\!&h_{12}^{-1}(2)h_{12}(1)v_2
\end{eqnarray}
and the received signals at $R_1$ and $R_2$ are given respectively
by
\begin{eqnarray}
Y_1(2)&\!\!\!=\!\!\!& h_{11}(1)v_1+h_{12}(1)v_2 \equiv
I_1(v_1,v_2)\\
Y_2(2)&\!\!\!=\!\!\!&h_{21}(2)h_{11}^{-1}(2)h_{11}(1)v_1+h_{22}(2)h_{12}^{-1}(2)h_{12}(1)v_2 \nonumber \\
&\!\!\!\equiv \!\!\!&L_2^2(v_1,v_2)
\end{eqnarray}
Hence, at the end of sub-phase one, $R_1$ has received
$I_1(v_1,v_2)$, the interference received in the first time
slot, and $R_2$ has received a new linear combination
$L_2^2(v_1,v_2)$.

In the second sub-phase, the transmitted signals are designed to
resurrect the interference received by $R_2$ in the first time
slot and provide a new linear combination of the desired messages
to $R_1$. The transmitted signals of $T_1$ and $T_2$ are given
respectively by
\begin{eqnarray}
X_1(3)&\!\!\!=\!\!\!&h_{21}^{-1}(3)h_{21}(1)u_1\\
X_2(3)&\!\!\!=\!\!\!&h_{22}^{-1}(3)h_{22}(1)u_2
\end{eqnarray}
where $T_1$ and $T_2$ utilize their perfect and instantaneous
knowledge of their channels to $R_2$.  The received signals at
$R_1$ and $R_2$ are given by:
\begin{eqnarray}
Y_1(3)&\!\!\!=\!\!\!&h_{11}(3)h_{21}^{-1}(3)h_{21}(1)u_1
+h_{12}(3)h_{22}^{-1}(3)h_{22}(1)u_2 \nonumber\\&\!\!\!\equiv
\!\!\!&
L_1^2(u_1,u_2) \\
Y_2(3)&\!\!\!=\!\!\!& h_{21}(1)u_1+h_{22}(1)u_2 \equiv
I_2(u_1,u_2)
\end{eqnarray}

After the third time slot, the two receivers $R_1$ and $R_2$ have
enough information to decode their intended messages.  In
particular, $R_1$ has access to two different equations in $u_1$
and $u_2$ only. The first one is obtained by subtracting $Y_1(2)$
from $Y_1(1)$ to cancel out the interference and the second
equation is $Y_1(3)$ by itself as it is received without
interference. Similarly, $R_2$ forms its first equation by
subtracting $Y_2(3)$ from $Y_2(1)$ to cancel out the interference
while the second equation is $Y_2(2)$.

Note that this scheme could be used also when the CSIT  pattern
given by (DD, NP, PN) but with minor modification in phase two,
where sub-phase one is dedicated to resurrecting interference of
$R_2$ instead of resurrecting interference of $R_1$ and sub-phase
two is dedicated to resurrecting interference of $R_1$ instead of
resurrecting the interference of $R_2$.

\subsection{Scheme 2: Distributed delayed and combined perfect CSIT}
Let us consider the 2-user SISO X-channel with alternating CSIT
given by (ND, DN, PP). Unlike case 1, here we have distributed
delayed CSIT over the first two time slots and combined perfect
CSIT in the last time slot. Consequently, the  interference
creation phase extends over two time slots while the interference
resurrection phase can be executed in one time slot as follows.

\textit{Phase one:} Each time slot of this phase is dedicated to
one receiver where the two transmitters transmit the desired
messages for this receiver. For example, if the first time slot is
dedicated to $R_1$, then $T_1$ transmits $u_1$ and $T_2$ transmits
$u_2$. The received signals at $R_1$ and $R_2$ are given
respectively by
\begin{eqnarray}
Y_1(1)&\!\!\!=\!\!\!&h_{11}(1)u_1 + h_{12}(1)u_2 \equiv
L_1^1(u_1,u_2)\\
Y_2(1)&\!\!\!=\!\!\!&h_{21}(1)u_1 + h_{22}(1)u_2 \equiv
I_2(u_1,u_2)
\end{eqnarray}
Therefore, $R_1$ receives linear combination $L_1^1(u_1,u_2)$ of
its desired signals, while $R_2$ receives only interference
$I_2(u_1,u_2)$. Similarly, in the next time slot, $T_1$
transmits $v_1$ and $T_2$ transmits $v_2$ and the received signals
at $R_1$ and $R_2$ are given respectively by
\begin{eqnarray}
Y_1(2)&\!\!\!=\!\!\!&h_{11}(2)v_1 + h_{12}(2)v_2 \equiv I_1(v_1,v_2)\\
Y_2(2)&\!\!\!=\!\!\!&h_{21}(2)v_1 + h_{22}(2)v_2 \equiv
L_2^1(v_1,v_2)
\end{eqnarray}
where  $R_2$ receives linear combination $L_2^1(v_1,v_2)$ of its desired signals, while $R_1$ receives only interference $I_1(v_1,v_2)$.

\textit{Phase two}: This phase includes only one time slot where the
transmitters resurrect the formerly received interference terms
$I_1(v_1,v_2)$ and $I_2(u_1,u_2)$, while providing new linear
combinations of the desired messages to the two receivers. In
order to achieve this goal, the transmitted signals from $T_1$ and
$T_2$ in the third time slot is given by
\begin{eqnarray}
X_1(3)&\!\!\!=\!\!\!&h_{21}^{-1}(3)h_{21}(1)u_1+h_{11}^{-1}(3)h_{11}(2)v_1\\
X_2(3)&\!\!\!=\!\!\!&h_{22}^{-1}(3)h_{22}(1)u_2+h_{12}^{-1}(3)h_{12}(2)v_2
\end{eqnarray}
and the corresponding received signals at $R_1$ and $R_2$ are
given respectively by
\begin{eqnarray}
Y_1(3)&\!\!\!=\!\!\!& L_1^2(u_1,u_2)+ I_1(v_1,v_2)\\
Y_2(3)&\!\!\!=\!\!\!& L_2^2(v_1,v_2)+ I_2(u_1,u_2)
\end{eqnarray}
where {\small
\begin{eqnarray}
 L_1^2(u_1,u_2)&\!\!\!\!=\!\!\!\!& h_{11}(3)h_{21}^{-1}(3)h_{21}(1)u_1
\!+\!h_{12}(3)h_{22}^{-1}(3)h_{22}(1)u_2 \qquad\\
L_2^2(v_1,v_2)&\!\!\!\!=\!\!\!\!&
h_{21}(3)h_{11}^{-1}(3)h_{11}(2)v_1\!+\!h_{22}(3)h_{12}^{-1}(3)h_{12}(2)v_2\qquad
\end{eqnarray} } \normalsize
At the end of the third time slot, each receiver can decode its
intended messages by solving two equations. For example, $R_1$
subtracts $Y_1(2)$ from $Y_1(3)$ to cancel out the interference
and obtain the first equation in $u_1$ and $u_2$ while the second
equation is $Y_1(1)$ by itself as it received without
interference.

Noteworthy, this scheme could be used when the CSIT pattern is
given by (DN, ND, PP) but with minor modification in phase one
where the two sub-phases swap their dedications from $R_1$ to
$R_2$ and vise versa.

\subsection{Scheme 3: Distributed delayed and distributed perfect CSIT}
As an illustrative example, let us consider a 2-user SISO X
channel with CSIT pattern given by (DN, PD, NP). Unlike the above
two examples, we have distributed delayed CSIT over the first two
time slots and distributed perfect CSIT over the last two
consecutive time slots. Consequently, the proposed scheme is
performed in two overlapping phases as follows.

\textit{Time slot 1:} The first sub phase of phase one begins at
$t=1$, and is dedicated to transmitting the desired messages of
$R_2$, i.e., $T_1$ transmits $v_1$ while $T_2$ transmits $v_2$.
The received signals are this given by
\begin{eqnarray}
Y_1(1)&\!\!\!=\!\!\!&h_{11}(1)v_1 + h_{12}(1)v_2 \equiv
I_1(v_1,v_2) \\
Y_2(1)&\!\!\!=\!\!\!&h_{21}(1)v_1 + h_{22}(1)v_2 \equiv
L_2^1(v_1,v_2)
\end{eqnarray}
Therefore, $R_2$ receives the first linear combination $L_2^1(v_1,v_2)$ of its desired signals, while $R_1$ receives only interference $I_1(v_1,v_2)$.

\textit{Time slot 2}: At $t=2$ the overlap occurs between the two
phases. In particular, sub-phase two of phase one and sub-phase
one of phase two begin simultaneously. In this time slot,
sub-phase two of phase one creates interference at $R_2$ with
while sub-phase one of phase two is designed to resurrect the
interference term $I_1(v_1,v_2)$. The transmitted signals are
given by:
\begin{eqnarray}
X_1(2)&\!\!\!=\!\!\!&u_1+h_{11}^{-1}(2)h_{11}(1)v_1 \\
X_2(2)&\!\!\!=\!\!\!&u_2+h_{12}^{-1}(2)h_{12}(1)v_2
\end{eqnarray}
and the corresponding received signals are given by:
\begin{eqnarray}
Y_1(2)&\!\!\!=\!\!\!&h_{11}(2)u_1 + h_{12}(2)u_2 +h_{11}(1)v_1 + h_{12}(1)v_2 \nonumber \\
&\!\!\!\equiv\!\!\!&  L_1^1(u_1,u_2)+I_1(v_1,v_2)\\
Y_2(2)&\!\!\!=\!\!\!& h_{21}(2)h_{11}^{-1}(2)h_{11}(1)v_1+h_{22}(2)h_{12}^{-1}(2)h_{12}(1)v_2 \nonumber \\
&\!\!\!+\!\!\!&h_{21}(2)u_1 + h_{22}(2)u_2\nonumber \\
&\!\!\!\equiv\!\!\!& L_2^2(v_1,v_2)+I_2(u_1,u_2)
\end{eqnarray}
Therefore, $R_2$ receives a new linear combination
$L_2^2(v_1,v_2)$ of its desired signals and an interference term
$I_2(u_1,u_2)$ as a by-product of the overlap, while $R_1$
receives the old interference $I_1(v_1,v_2)$ and the first linear
combination $L_1(u_1,u_2)$ of its desired signals.

\textit{Time slot 3}:  In this time slot the  transmitters send
linear combination from $u_1$ and $u_2$ aiming to resurrect the
interference terms $I_2(u_1,u_2)$ formerly received at $t=2$,
while providing a new linear combinations to $R_1$ of its desired
messages. The transmitted signals are given by:
\begin{eqnarray}
X_1(3)&\!\!\!=\!\!\!&h_{21}^{-1}(3)h_{21}(2)u_1\\
X_2(3)&\!\!\!=\!\!\!&h_{22}^{-1}(3)h_{22}(2)u_2
\end{eqnarray}
and the corresponding received signals are
\begin{eqnarray}
Y_1(3)&\!\!\!=\!\!\!&h_{11}(3)h_{21}^{-1}(3)h_{21}(2)u_1
+h_{12}(3)h_{22}^{-1}(3)h_{22}(2)u_2\nonumber \\
&\!\!\!\equiv\!\!\!& L_1^2(u_1,u_2)\\
Y_2(3)&\!\!\!=\!\!\!&h_{21}(2)u_1+h_{22}(2)u_2 \equiv
I_2(u_1,u_2)
\end{eqnarray}
Finally, the two receivers $R_1$ and $R_2$ have enough information
to decode their intended messages.  In particular, $R_1$ has
access to two different equations in $u_1$ and $u_2$ only. The
first one is obtained by subtracting $Y_1(1)$ from $Y_1(2)$ to
cancel out the interference and the second equation is $Y_1(3)$ by
itself as it is received without interference. Similarly, $R_2$
its first equation is $Y_2(1)$ while forming its second equation
by subtracting $Y_2(3)$ from $Y_2(2)$ to cancel out the
interference.

Note that this scheme could be used when the CSIT pattern is given
by (ND, DP, PN)  but with minor modification in the two phases
where the two sub-phases in each phase swap their dedications from
$R_1$ to $R_2$ and vise versa .
\vspace{5 mm}
\section{Synergistic CSIT Alternation Patterns}
In this section, we discuss CSIT alternation patterns that can
provide synergistic gain in the DoF of the two-user SISO X
channel. We focus on a three-symbol extension of the channel.
Since the possible CSIT states for the two users are given by
$S_{12}\in \{\text{PP,PD,PN,DP,DD,DN,NP,ND,NN}\}$, there are $9^3$
possible alternation patterns over the three time slots.

First, we note that the aforementioned three examples in Section
\ref{section_schems} present the CSIT patterns with the lowest
CSIT sufficient to achieve $4/3$ DoF. Definitely, any alternation
pattern with channel knowledge higher than these patterns can
achieve the same DoF, i.e., if we have $S_{12}= \text{ND}$, its
higher CSIT state that could provide the same synergistic DoF gain
are $\{\text{NP, DD, DP, PD, PP}\}$. Theorem 1 presents sufficient
conditions on the lowest CSIT alternation pattern among tree-symbol channel extension patterns for achieving
the upper bound on the DoF of the two-user SISO X-channel.\\

\textit{Theorem 1:} For the two-user SISO X-channel in time
varying or frequency selective settings, the upper bound on the
total DoF of the channel is achievable if the following 
requirements on the CSIT alternation pattern are satisfied.
\begin{enumerate}
    \item Each transmitter has a delayed CSIT followed by a perfect CSIT over three time slots.
    \item At each time slot, at least one transmitter should have some CSIT (perfect or delayed),
    i.e., the two transmitters should not be simultaneously
    without CSIT.
    \item In the third time slot, at least one transmitter should
    have perfect CSIT.\\
\end{enumerate}

\textit{Proof:} We show that the three requirements of Theorem 1
limit the CSIT alternation patterns in
 a 3-symbol channel extension to the minimum CSIT
synergistic patterns considered in the three examples of Section
\ref{section_schems}, and its higher CSIT patterns. Hence, the
achievability of $4/3$ DoF follows from the results of Section
\ref{section_schems}. The first requirement in Theorem 1 yields
three possible minimum states for the CSIT of the channel to the
$i$th receiver over three time slots $S_i^T \in \{\text{(D,P,N),
(D,N,P), (N,D,P)}\}$. As a result we have 9 possible combinations
for the CSIT of the two-user channel. Six of these 9 combinations,
satisfy the second and third requirements in Theorem 1 and are
listed as the first 6 entries in Table 1. The remaining three
combinations are those which have $S_{12}= \text{NN}$ in any of
the three time slots, i.e., (DD,PP,NN), (DD,NN,PP), and
(NN,DD,PP). For the first combination, the minimum CSIT states
that satisfy the three requirements are (DD,PP,PN), which is
higher than (DD,NP,PN), and (DD,PP,NP), which is higher than
(DD,PN,NP). From Table 1, we can achieve $4/3$ DoF using
achievable scheme 1 in both cases. Similarly, for the CSIT state
(DD,NN,PP), the minimum CSIT that satisfy the requirements of
Theorem 1 are (DD,ND,PP) and (DD,DN,PP) for which $4/3$ DoF can be
achieved using scheme 2. Finally, for the CSIT state (NN,DD,PP),
the minimum CSIT that satisfy the requirements of Theorem 1 are
(ND,DD,PP) and (DN,DD,PP) for which $4/3$ DoF can be achieved
using scheme 2
too. \\
\vspace{-3mm}
\begin{table}[!ht]
\begin{tabular}{|p{1.8cm}|p{1.7cm}||p{1.8cm}|p{1.7cm}|}
  \hline
   CSIT state & Scheme &CSIT state &  Scheme  \\
  \hline
 (DD,PN,NP) & Scheme 1 &(DN,PD,NP) & Scheme 3\\
 (DD,NP,PN) & Scheme 1 &(DN,ND,PP) & Scheme 2\\
 (ND,DP,PN) & Scheme 3 &(ND,DN,PP) & Scheme 2\\
  \hline
\end{tabular}\\
\caption {Achievable schemes for different CSIT states}
\end{table}

\textit{Remark 1:} \textbf{[Synergy benefits]} Note that the DoF for 
two-user X-channel with perfect CSIT is
$4/3$ \cite{maddah2008communication, 4418479}, with delayed CSIT
is upper bounded by $6/5$ \cite{Delayed_CSIT} and with No-CSIT is
unity due to statistical symmetry of channel outputs \cite{6205390}. Synergy is the interaction
of multiple elements in a system to produce an effect greater than
the sum of their individual effects. In particular , the
alternation of CSIT states $S_{ij}$ over three time slots works
cooperatively to provide a DoF greater than the DoF of the weighted average of
their individual DoF for the same network.

As an example, let us consider the CSIT alternation pattern given
by (DD,DD,PP). If there is no interaction between the three time
slots, the DoF that can be obtained are given by $\frac{2}{3}
\frac{6}{5} + \frac{1}{3} \frac{4}{3}= \frac{56}{45}$ which is strictly
lower than the upperbound on the DoF of the channel. However,
using achievable scheme 2, we can get 4/3 DoF for this case as
this CSIT pattern is higher than (DN,ND,PP) in Table 1. This
illustrates the synergistic benefit that can be obtained from the
alternation of CSIT over the three time slots.

Note that not all combinations of CSIT states could provide
synergistic benefits or work together in a cooperative way. For
example, when perfect CSIT comes before delayed and no CSIT , it
loses its synergetic DoF gain; where, the DoF degrades to the sum
of the individual DoF each case. on the other hand, when perfect
CSIT comes after delayed and no CSIT, the synergy of the alternation appears.\\

\textit{Remark 2:} \textbf{[Potential of delayed followed by
perfect CSIT]} The extraordinary synergistic gain of delayed CSIT
followed by perfect CSIT lies in the ability to upgrade the X
channel to a broadcast channel with delayed CSIT. In particular,
when the delayed CSIT comes first it provides the transmitters
with delayed channel knowledge which combats the distributed
nature of the X-channel and can be exploited in addition to the
perfect CSIT to provide one message to each receiver.\\

\textit{Remark 3:} \textbf{[Combined No-CSIT]} We note that the
synergy of alternating CSIT is lost when the network has combined
No-CSIT in any time slot. Intuitively, the uncertainty of
distributed channel unawareness is better than blindness of
combined complete ignorance. As Martin Luther king said before
``Darkness cannot drive out darkness; only light can do that.''

\section{Conclusion}
We have investigated the synergistic benefits of alternating CSIT
in the context of two-user X-channel.
Unlike what is commonly thought that the synergistic benefits of
alternating CSIT could be more sensitive to or may be lost
depending on whether the transmitter of the network are
distributed or collocated, we end up with the synergistic
alternation of CSIT is still very beneficial in distributed
transmitters network.
The surprising finding of
the synergistic alternating CSIT is that it is capable of achieving the information theoretic upper bound upper bound on the DoF of the two-user SISO X-channel. Hence, \textit{constantly} perfect CSIT has much redundant and unnecessary channel information.

\bibliographystyle
{IEEEtran}
\bibliography{IEEEabrv,Nulls}

\begin{thebibliography}{10}
\providecommand{\url}[1]{#1}
\csname url@samestyle\endcsname
\providecommand{\newblock}{\relax}
\providecommand{\bibinfo}[2]{#2}
\providecommand{\BIBentrySTDinterwordspacing}{\spaceskip=0pt\relax}
\providecommand{\BIBentryALTinterwordstretchfactor}{4}
\providecommand{\BIBentryALTinterwordspacing}{\spaceskip=\fontdimen2\font plus
\BIBentryALTinterwordstretchfactor\fontdimen3\font minus
  \fontdimen4\font\relax}
\providecommand{\BIBforeignlanguage}[2]{{%
\expandafter\ifx\csname l@#1\endcsname\relax
\typeout{** WARNING: IEEEtran.bst: No hyphenation pattern has been}%
\typeout{** loaded for the language `#1'. Using the pattern for}%
\typeout{** the default language instead.}%
\else
\language=\csname l@#1\endcsname
\fi
#2}}
\providecommand{\BIBdecl}{\relax}
\BIBdecl

\bibitem{jafar2011interference}
S.~A. Jafar, \emph{Interference alignment: A new look at signal dimensions in a
  communication network}.\hskip 1em plus 0.5em minus 0.4em\relax Now Publishers
  Inc, 2011.

\bibitem{maddah2008communication}
M.~A. Maddah-Ali, A.~S. Motahari, and A.~K. Khandani, ``Communication over
  {MIMO} {X} channels: Interference alignment, decomposition, and performance
  analysis,'' \emph{IEEE Transactions on Information Theory}, vol.~54, no.~8,
  pp. 3457--3470, 2008.

\bibitem{4567443}
V.~Cadambe and S.~Jafar, ``Interference alignment and degrees of freedom of the
  {K}-user interference channel,'' \emph{IEEE Transactions on Information
  Theory}, vol.~54, no.~8, pp. 3425--3441, 2008.

\bibitem{5208535}
------, ``Interference alignment and the degrees of freedom of wireless {X}
  networks,'' \emph{IEEE Transactions on Information Theory}, vol.~55, no.~9,
  pp. 3893--3908, 2009.

\bibitem{4418479}
S.~Jafar and S.~Shamai, ``Degrees of freedom region of the {MIMO} {X}
  channel,'' \emph{IEEE Transactions on Information Theory}, vol.~54, no.~1,
  pp. 151--170, 2008.

\bibitem{6205390}
C.~Vaze and M.~Varanasi, ``The degree-of-freedom regions of {MIMO} broadcast,
  interference, and cognitive radio channels with no {CSIT},'' \emph{IEEE
  Transactions on Information Theory}, vol.~58, no.~8, pp. 5354--5374, 2012.

\bibitem{maddah2012completely}
M.~A. Maddah-Ali and D.~Tse, ``Completely stale transmitter channel state
  information is still very useful,'' \emph{IEEE Transactions on Information
  Theory}, vol.~58, no.~7, pp. 4418--4431, 2012.

\bibitem{maleki2012retrospective}
H.~Maleki, S.~A. Jafar, and S.~Shamai, ``Retrospective interference alignment
  over interference networks,'' \emph{IEEE Journal of Selected Topics in Signal
  Processing}, vol.~6, no.~3, pp. 228--240, 2012.

\bibitem{yang2012degrees}
S.~Yang, M.~Kobayashi, D.~Gesbert, and X.~Yi, ``Degrees of freedom of {MISO}
  broadcast channel with perfect delayed and imperfect current {CSIT},'' in
  \emph{Information Theory Workshop, (ITW) 2012}.

\bibitem{6471826}
R.~Tandon, S.~Jafar, S.~Shamai~Shitz, and H.~Poor, ``On the synergistic
  benefits of alternating csit for the miso broadcast channel,'' \emph{IEEE
  Transactions on Information Theory}, vol.~59, no.~7, pp. 4106--4128, 2013.

\bibitem{ghasemi2011degrees}
A.~Ghasemi, A.~S. Motahari, and A.~K. Khandani, ``On the degrees of freedom of
  {X} channel with delayed {CSIT},'' in \emph{proceedings IEEE International
  Symposium on Information Theory (ISIT) 2011}.

\bibitem{Delayed_CSIT}
S.~Lashgari, A.~S. Avestimehr, and C.~Suh, ``Linear degrees of freedom of the
  x-channel with delayed csit,'' \emph{CoRR}, vol. abs/1309.0799, 2013.

\end{thebibliography}

\end{document}